\documentclass{kapproc} % Computer Modern font calls

\RequirePackage{graphicx}%
\RequirePackage{epsf}%
\input{psfig.sty}

\def\Lsun{\hbox{\it L$_\odot$}}

\def\Msun{\hbox{\it M$_\odot$}}

\def\Msunyr{\hbox{\it M$_\odot\,$yr$^{-1}$}}
\def\Myr{\hbox{\it Myr}}

\def\simgr{\mathrel{\hbox{\rlap{\hbox{\lower4pt\hbox{$\sim$}}}\hbox{$>$}}}}
\def\apj{\hbox{ApJ}}
\def\apjl{\hbox{ApJL}}
\def\aj{\hbox{AJ}}
\def\aap{\hbox{A\&A}}
\def\aaps{\hbox{A\&AS}}
\def\mnras{\hbox{MNRAS}}
\def\araa{\hbox{ARA\&A}}
\def\nat{\hbox{Nature}}

\upperandlowercase
\let\footnote\savefootnote
\let\footnotetext\savefootnotetext 
 
\kluwerbib % will produce  bibliography 
\begin{document}

%------------ article title  ------------------->>

\articletitle{The Stellar Initial Mass Function in the Galactic Center}

%% optional, to supply a subtitle:
%\articlesubtitle{Spineto@50}

%% Supply a shorter version of the title for the running head:
\chaptitlerunninghead{IMF in GC}

%------ author/affiliation choices -------------->>

%% Single author or several authors with same affiliation

 \author{Donald F. Figer}
 \affil{STScI, 3700 San Martin Drive, Baltimore, MD 21218, USA}
 \email{figer@stsci.edu}

%% Multiple authors, multiple affiliations

%\author{First Author\altaffilmark{1}, Second Author\altaffilmark{2}, 
%         Third Author\altaffilmark{1,3}}

%\affil{\altaffilmark{1}Institute, Address, Country, \\ 
%\altaffilmark{2}Institute, Address, Country, \\
%\altaffilmark{3}Institute, Address, Country}

%\email{author1@add1,author2@add2,author3@add3}

% abstract
 \begin{abstract}
Massive stars define the upper limits of the star formation process, dominate 
the energetics of their local environs, and significantly affect the chemical 
evolution of galaxies. Their role in starburst galaxies and the early Universe 
is likely to be important, but we still do not know the maximum mass that a 
star can possess, i.e.``the upper mass cutoff.'' I will discuss results from 
a program to measure the upper mass cutoff and IMF slope in the Galactic Center. 
The results suggest that the IMF in the Galactic center may deviate significantly 
from the Salpeter value, and that there may be an upper mass cutoff to the 
initial mass function of $\sim$150 Msun. 
 \end{abstract}

%------------ body of article ------------------->>
\section{Motivating Questions}
Two simple, yet still unanswered, questions motivate this paper. First, is the
stellar initial mass function (IMF) universal? Second, what is the most massive
star that can form? These questions are related, as they concern 
primary output products of the star formation process. The IMF is observed to be roughly constant, within errors, 
although outliers to the value of the slope do exist. The data at the high mass end are woefully incomplete
for determining the upper limit for which the IMF essentially becomes zero, i.e.\ an
upper mass cutoff. 

There are several properties of stellar clusters that are required 
for estimating the high mass IMF slope and, in particular, an upper mass cutoff: 
\begin{enumerate}
\item
the associated star formation event must produce a large
amount of mass in stars, at least 10$^4$~\Msun, 
\item
the resultant cluster must be young enough,
certainly no older than 3~\Myr, so that its most massive members are pre-supernovae, 
\item
the cluster must be old enough for its stars to have
emerged from their natal cocoons, 
\item
the cluster must be close enough to be resolved into
individual stars, and 
\item 
the stellar surface number density must be low enough to allow
one to separate light from individual stars. 
\end{enumerate}
Given this rather long list of requirements, it is
perhaps not surprising that, as of yet, an upper mass cutoff has not been identified; although, 
recent work might have identified a cutoff in R136, a starburst cluster in
the LMC (\cite{wei04}). 
There is only one cluster in the Galaxy that can satisfy these requirements, the Arches
cluster near the Galactic center. There are two other clusters massive enough, the Quintuplet
and Central clusters, also both in the Galactic center, but those clusters are both too
old, $\sim$4~\Myr, and their most massive members have dimmed as WR stars or compact
objects.  

\section{The Galactic center environment and its young clusters}
The Galactic center occupies a very small volume, $\sim$0.04\%\ of the Galaxy, yet
it contains 10\%\ of all Galactic molecular material and a similar proportion
of newly formed stars. The extreme tidal forces in the center shred molecular
clouds having densities less than about 10$^{4}$~cm$^{-3}$; therefore, the
clouds in the region necessarily have relatively high densities compared to those in the disk.
The cloud temperatures are also about a factor of three higher, and the
magnetic field strength may be as much as 1~mG. 
This environment may favor the formation of massive stars (\cite{mor93}). 
See Morris \& Serabyn (1996) for a review.

There are three massive young clusters within a projected radius
of 30~pc of the Galactic center: the Arches, the Quintuplet, and the Central cluster. 
Their properties have been reviewed 
(\cite{fig99a,fig99c,fig02,fig03}). In brief, all three have about equal mass, $\sim$10$^4$~\Msun; but
the first is only 2-2.5~\Myr\ old, or about half the age of the other two. These older 
clusters are too old for an accurate determination of their initial mass functions using
photometry alone, as their most massive members have likely progressed to the supernovae
stage, or, at the least, have dimmed substantially and are lost amongst the background
population of red giants. Even when they are distinguished from the background, as Wolf-Rayet
stars, it is impossible to infer their original masses. 

\section{The Arches cluster and the IMF in the Galactic center}
The Arches cluster is located just 30~pc, in projection, from the Galactic 
center (\cite{cot92,nag95,fig95a,cot95,cot96,blu01}).
It contains 160~O-stars, and is the most massive young cluster in the Galaxy (\cite{ser98,fig99a}). 
Given its youth, its members have not yet advanced
to their end states, and they still follow a linear relationship between mass and magnitude (\cite{fig02}). 
The brightest members have masses $\sim$120~\Msun, and are
enriched in helium and nitrogen (\cite{naj04}). They are commensurately luminous, up to 10$^{6.3}$~\Lsun,
and have prodigious winds that carry a significant amount of mass, up to 10$^{-5}$~\Msunyr. Some of
these winds have been individually identified through radio observations of their free-free
emission (\cite{lan01}), and there are several point-like and diffuse x-ray emission sources
centered on the cluster (\cite{yus02}). 

The IMF in the Galactic center has been measured in one location, the Arches cluster (\cite{fig99a,sto02}), and
it is somewhat shallow compared to the Salpeter value (\cite{sal55}). Figure~1 shows the mass function,
as observed using HST/NICMOS. It was constructed by converting magnitudes into masses using the
Geneva stellar evolution models (\cite{sch92}). The counts have been corrected for contamination by the 
background population. The slope appears to be shallow with respect to the Salpeter value. 
Such a cluster is likely to have experienced significant dynamical evolution in the strong
tidal field of the Galactic center; however, Kim et al.\ (2000) find that this effect is unlikely
to have produced such a flat slope.  

\begin{figure}[ht]
\centerline{\psfig{file=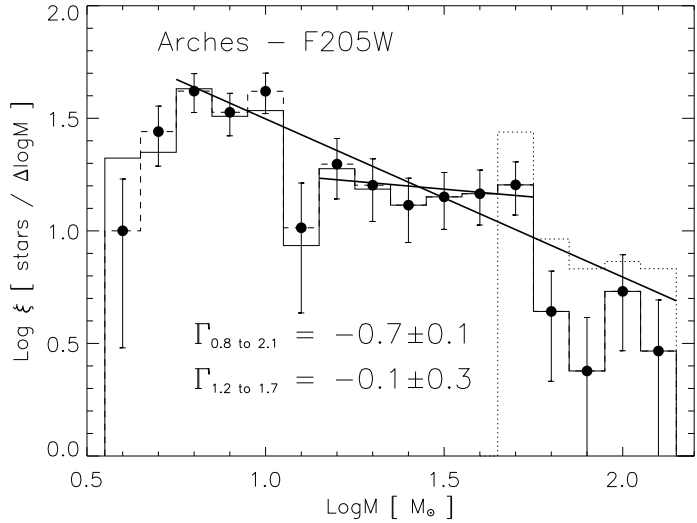,height=3in,angle=90}}
\caption{Present-day mass function of the Arches cluster in the F205W NICMOS filter (bold). Incompleteness
corrected data are shown with a dashed line. The dotted line shows concurrence with
earlier observations using Keck for the highest mass stars. The average IMF slope is $-$0.7, although
Stolte et al. (2002) found a slightly steeper slope of $-$0.9 after correcting for
differential extinction.}
\end{figure}

\section{An upper mass cutoff}
The Arches cluster appears to have an upper mass cutoff (\cite{fig03,fig04,fig05}). 
Figure~2 shows the mass function extended to very high masses and computed out to a radius
of 12 arcseconds from the center of the Arches cluster. In this plot, we see that one might
expect massive stars up to 500$-$1,000~\Msun, yet none are seen beyond $\sim$120~\Msun.
Taking account of errors, and the unreliable mass-magnitude relation at the highest
masses, one can safely estimate an upper mass cutoff of $\sim$150~\Msun. The most 
important caveat to this result relates to the youth of the cluster. An age $>$3~\Myr\ would
mean that the most massive stars have progressed to their end states and would not
be observed. Several analyses suggest an age of 2-2.5~\Myr\ (\cite{fig99a,blu01,fig02,naj04}).
Note that an age $<$1~\Myr\ would give a deficit of roughly twice that shown in the
figure and a predicted maximum stellar mass of 600$-$1,700~\Msun.
A similar analysis was done for R136 in the LMC that also found a cutoff of 150~\Msun\ (\cite{wei04}).

\begin{figure}[ht]
\centerline{\psfig{file=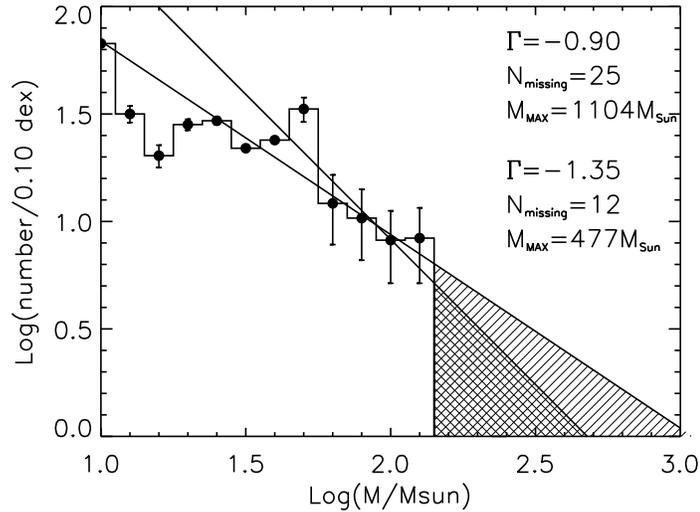,height=3in,angle=90}}
\caption{Present-day mass function of the Arches cluster in the F205W NICMOS filter with
lines overplotted for the inferred mass function and the Salpter mass function. The hatched
regions demonstrate that one would expect there to be many very massive stars in the
cluster (Figer 2005).}
\end{figure}

\section{Supermassive stars in violation of the cutoff?}
There are several stars in apparent violation of the cutoff estimated in the Arches
cluster. The Pistol star, in the Quintuplet cluster, has an estimated initial mass of 150$-$200~\Msun\ (\cite{fig98,fig99b}).
Star \#362 in the Quintuplet cluster is a near-twin to the Pistol star (\cite{fig99c,geb00}), so it
likely had a similar initial mass. Both stars are roughly equally bright, although 
they are also both variable (\cite{gla01}). It is interesting to note that these two stars,
if as massive as we think, should only live for $\sim$3~\Myr, yet they reside in a cluster
that is at least 4~\Myr\ old. 
LBV1806$-$20 is another high mass star that may violate the limit, but it is
likely binary (\cite{fig04b}). $\eta$ Car, with a system mass that may be higher than
the limit, is also a likely binary (\cite{dam00}). A promising resolution to the apparent
paradox of a limit and systems with higher masses could be that all such systems are either
binary or have been built through recent mergers (\cite{fig02x}). 

%\begin{acknowledgments}
%....Insert acknowledgments...
%\end{acknowledgments}

\end{document}